\documentclass[a4paper,12pt]{article}
\usepackage[utf8]{inputenc}
 \usepackage[margin=1.1in]{geometry}
 \usepackage{amssymb} 
\usepackage{amsmath}
\usepackage{authblk}
\usepackage{comment}
\usepackage{tikz}
\usepackage{graphicx}

\usepackage{slashed}

\usepackage{multicol}

\usepackage{comment}

\usepackage[hyperindex,breaklinks]{hyperref}
\usepackage[numbers,sort&compress]{natbib}

\title{\Large Vacuum Stability vs. Positivity in Real Singlet Scalar Extension of the Standard Model}
\author{\vspace*{2cm}Parsa Ghorbani}
\affil{\normalsize \it Dipartimento di Fisica dell'Universit{\`a} di Pisa, Italy \\
\normalsize \it INFN, Sezione di Pisa, Italy\\

}
\date{}

\begin{document}
\maketitle

\begin{abstract}
We assume a generic real singlet scalar extension of the Standard Model living in the vacuum $(v,w)$ at the electroweak scale with $v=246$ GeV and $w$ being respectively the Higgs and the singlet scalar vacuum expectation values. By requiring {\it absolute} vacuum stability for the vacuum $(v,w)$, the positivity condition and the perturbativity up to the Planck scale, we show that the viable space of parameters in the model is strongly constrained for various singlet scalar vacuum expectation values $w=0.1, 1, 10, 100$ TeV. Also, it turns out that the singlet scalar mass can be from a few GeV up to less than TeV.
\end{abstract}

\newpage

\section{Introduction}
\label{int}
The stability of the vacuum in the Standard Model (SM) was first utilized as an implement to put a bound on the Higgs mass and the mass of fermions in the SM framework\cite{Cabibbo:1979ay}. After the discovery of the Higgs particle by ATLAS and CMS experiments at the LHC in 2012, the value of the Higgs mass is determined accurately to be around $125$ GeV \cite{Aad:2012tfa,Chatrchyan:2012ufa}. The mass of the top quark (as the heaviest quark) was already known to be about $176$ GeV \cite{Abe:1995hr}. Having the Higgs and the top quark masses in hand, and knowing the value of the Higgs vacuum expectation value (VEV) to be $v=246$ GeV, the status of the vacuum stability in the SM becomes lucid; the SM vacuum starts to be metastable at energy scale around $10^{10}$ GeV \cite{Bezrukov:2012sa,Degrassi:2012ry,Buttazzo:2013uya}. This happens because in the SM, the top quark has a large negative contribution in the renormalization group equations (RGE) for the Higgs quartic coupling $\lambda_\text{h}$, so that the Higgs quartic coupling becomes negative at higher energy scales which makes the vacuum with $v\neq 0$ metastable.  

There is a consensus in the literature that in the presence of more scalars, the vacuum can become stable up to the Planck scale. For instance, the vacuum stability in extensions of the SM by adding an extra real scalar with $\mathbb{Z}_2$ symmetry is studied in \cite{Gonderinger:2009jp,Falkowski:2015iwa}, employing a complex scalar to address the vacumm stability in \cite{EliasMiro:2012ay,Gonderinger:2012rd}, using scalars in scale invariant extension of the SM in \cite{Gabrielli:2013hma, Khoze:2014xha}, investigation of the di-Higgs production in singlet scalar model in \cite{Chen:2014ask}, studying the vacuum stability of 2HDM at the electroweak (EW) scale in \cite{Ferreira:2004yd}, and stablizing the vacuum by Higgs–inflaton mixing in \cite{Ema:2017ckf}. 

The point that we want to emphasize in this paper is the importance of the positivity condition (i.e. the requirement of having a positive definite potential in all scales), when studying the vacuum stability in a given model. At the EW scale, say at the scale $\mathcal{O}(m_t)$, the free parameters of an extended SM model must be chosen in a way to respect the positivity condition. However, when solving the RGEs there is no guarantee that the positivity condition will be satisfied in higher energy scales. This should be considered alongside the possible change of the vacuum structure at higher scales. 

In the case of the SM, the vacuum stability and the positivity are delicately related. The potential in the SM is given by $V_\text{SM}=-\mu_\text{h} h^2/2+ \lambda_\text{h} h^4/4$ for which if $\lambda_\text{h}>0$ the theory develops a non-zero VEV for the Higgs scalar. If $\lambda_\text{h}<0$ the Higgs VEV can only be vanishing. Due to the large negative contribution of the heavy top quark in the RGE for  $\lambda_\text{h}$, the Higgs quartic coupling becomes negative at the scale $10^{10}$ and thereafter the Higgs non-zero VEV is no longer a minimum of the theory. Therefore, in the SM the sign of the quartic coupling $\lambda_\text{h}$ changes the structure of the vacuum and deals with the vacuum stability. 
At the same time, the sign of the Higgs quartic coupling confirms or violates the positivity of the potential. If $\lambda_\text{h}<0$ the theory is no longer well defined before thinking about the vacuum structure of the model. Therefore, in the case of the SM, the quartic coupling $\lambda_\text{h}$ plays a dual role as a tuner for the vacuum stability and the positivity of the model. 

When more scalars are added in the theory, the vacuum stability and the positivity condition should be investigated  separately. Both conditions even if consistent at the EW scale, may become in conflict at higher scales. This might resemble although maybe not directly related to the situation that for a random set of parameters in a multi-scalar potential having at the same time a small Higgs mass and a small cosmological constant is not very probable (see \cite{Ghorbani:2019zic}). 
In general with more scalars the vacuum structure of the model gets more complicated; as the number of vacua grows rapidly with the number of scalars (see \cite{Ghorbani:2019itr} for two-scalar example). The positivity condition which is obtained from the quartic part of the potential, can be very involved in general with more scalars \cite{Kannike:2016fmd}, unless some symmetry is applied. 

The perturbativity of the theory is another constraint which must be taken into account when running the parameters of the model to higher scales. The main question is how consistent the vacuum stability constraint, the positivity condition and the perturbativity  are in a multi-scalar theory from the electroweak up to the Planck scale.   

In this article, as the first step to answer the question posed above, we extend the SM with a generic real scalar potential including also the scalar cubic term and linear scalar-Higgs interaction. We will study the scale evolution of the only vacuum of the model i.e. $(v,w)$ at the EW scale, and will argue the {\it absolute} vacuum stability up to the Planck scale. By absolute vacuum stability we mean that the potential possess only one single minimum for whole range of the energy scale here from the EW up to the Planck.  This will be confronted with the scale evolution of the positivity condition as well as the perturbativity up to the Planck scale. Using the aformentioned constraints we put strong bound on free parameters of the model. For simplicity we use the term {\it SPP conditions} when we consider the stability, positivity and perturbativity conditions altogether.  

The rest of the paper is arranged as the following. In section \ref{model} we introduce the model giving the details of the vacuum solution $(v,w)$ and the positivity condition. In section \ref{rge} we will discuss the RGEs, and will present our numerical analysis in section \ref{numeric}. We will bring a summery of the results in section \ref{concl}.

\section{Vacua in Singlet Scalar Model}\label{model}
The vacuum stability in the real singlet scalar extension of the SM with $\mathbb{Z}_2$ symmetry has been studied in \cite{Falkowski:2015iwa}. The presence of the $\mathbb{Z}_2$ symmetry simplifies the model considerably. For such model provided that the positivity condition is taken into account, if the vacuum solution $(v,w)$ is a minimum at a given scale $\Lambda$, it remains the global minimum in all scales because $(v,w)$ and other extremum solutions of the $\mathbb{Z}_2$ symmetric model i.e. $(0,0)$, $(v,0)$ and $(0,w)$ cannot be minimum at the same time (see \cite{Ghorbani:2020xqv} on vacuum structure of the $\mathbb{Z}_2$ symmetric singlet scalar model). 

Here we consider instead a generic real scalar extension of the SM without the $\mathbb{Z}_2$ symmetry,

\begin{table}
 \begin{center}
   \small \begin{tabular}{ | l | l | l  |l| l| l| l| p{2cm} |}
    \hline
    $w$ & $100$ GeV&$1$ TeV &$10$ TeV&$100 $ TeV\\ \hline 
 $\lambda_\text{h}$    & $(0.39,0.51)$ & $(0.1,0.54)$ & $(1.9\times 10^{-4},0.39)$  & $(0.95,1)$  \\ \hline
   $\lambda_\text{s}$ & $(0.25,1)$ & $(0.015 , 0.032)$& $(2.4,4.07)\times 10^{-4}$ & $(1.03,1.13)\times 10^{-5}$  \\ \hline
     $\lambda_\text{hs}$ &$(-0.17,-0.02)$  &$(-0.033,0.015)$ & $(-2900,-4.53)\times 10^{-6}$ &$(-2.37,-0.96)\times 10^{-6}$  \\ \hline
     $\kappa_\text{s}$  & $(-1,1)$ &$(-1,1)$ & $(-1,1)$ &$(0.49,0.64)$\\ \hline
      $\kappa_\text{hs}$  & $ (-1,1)$& $(-0.054,1)$& $(-1,0.097)$ & $(-0.55,-0.098)$\\ \hline
      $m_s$ & $191$-$257$ GeV  & $218$-$308$ GeV & GeV $185$-$334$ GeV& $553$-$574$ GeV \\ \hline
      \end{tabular}
\end{center}
        \caption{The allowed region for the couplings $\lambda_\text{h},\lambda_\text{s},\lambda_\text{hs}$, $\kappa_\text{s},\kappa_\text{hs}$ and $m_s$ for different singlet scalar VEV benchmarks $w=0.1, 1, 10, 100$ TeV, respecting the absolute vacuum stability and positivity condition for the vacuum $(v,w)$ at the electroweak scale with the assumptions $m_H=125$ GeV, $v_H=246$ GeV and $m_s>m_H$.}
        \label{PTtable1}
        \end{table} 

\begin{equation}\label{sinpot}
 V(h,s)=-\frac{1}{2}\mu^2_\text{h} h^2 + \frac{1}{4}\lambda_\text{h} h^4  -\frac{1}{2}\mu^2_\text{s} s^2 +\frac{1}{3} \kappa_\text{s} s^3 + \frac{1}{4}\lambda_\text{s} s^4
 +\frac{1}{2} \kappa_\text{hs} h^2 s +\frac{1}{4} \lambda_\text{hs} h^2 s^2 \,.
 \end{equation}
The positivity condition on the quartic part of the potential above is the same as the $\mathbb{Z}_2$ symmetric potential and is given by $\lambda_\text{h}>0$, $\lambda_\text{h}>0$ and
\begin{equation}\label{pos}
  \lambda_\text{hs} > 0  ~\vee~ \left( \lambda_\text{hs}<0  ~\wedge~ \lambda^2_\text{hs}\leq \lambda_\text{h} \lambda_\text{s}\right) \,. 
\end{equation}
The vacuum solutions for the potential in Eq. (\ref{sinpot}) can only have the structures $(0,0)$, $(0,w)$ or $(v,w)$; the VEV solution $(v,0)$ is not allowed. Despite the $\mathbb{Z}_2$ symmetric model ( i.e. the potential in Eq. (\ref{sinpot}) with $\kappa_\text{hs}=\kappa_\text{s}=0$), all three extremum solutions can be local minimum at the same time even if we take into account the positivity condition. 
For the $\mathbb{Z}_2$ symmetric model after the electroweak symmetry breaking the possible minima at the EW scale are either $(v,w)$ or $(v,0)$, but for the generic potential in Eq. (\ref{sinpot}) the only vacuum solution at the EW scale is inevitably $(v,w)$ 
which is given by,
\begin{equation}
 v= \frac{\sqrt{\mu^2_\text{h}-\kappa_\text{hs} w -\lambda_\text{hs} w^2}}{\sqrt{\lambda_\text{h}}}, \hspace{2cm} w= p+ \xi_-^{1/3}+ \xi+^{1/3},
\end{equation}
where
\begin{equation}\label{vwsol}
\begin{split}
 &\xi_\pm=  q\pm\sqrt{q^2+\left(r-p^2 \right)^3},  \\
%  &p=\frac{2\kappa_\text{s}\lambda_\text{h}-3\kappa_\text{hs}\lambda_\text{hs}}{6(\lambda^2_\text{hs}-\lambda_\text{h} \lambda_\text{s})},  \\
 &q=\frac{b^3 +9a (6 \kappa_\text{hs} \mu^2_\text{h} a - b c) }{216 a^3},~~~~~~p=\frac{b}{6a},~~~~~~~r=\frac{c}{6a}\,,\\
  &a=\lambda^2_\text{hs}-\lambda_\text{h} \lambda_\text{s},~~~~~
 b=2\kappa_\text{s} \lambda_\text{h}-3 \kappa_\text{hs}\lambda_\text{hs},~~~~~
 c=\kappa^2_\text{hs}-2\lambda_\text{hs} \mu^2_\text{h}+2\lambda_\text{h} \mu^2_\text{s}\,.
  \end{split}
\end{equation}
From Eq. (\ref{vwsol}) real solutions for $v$ and $w$ requires $q^2>(p^2-r)^{1/3}$ for the reality of $w$ and $\mu^2_\text{h}>0, \lambda_\text{hs}\leq -\kappa^2_\text{hs}/4\mu^2_\text{h} $ for the reality of $v$.        
The parameters $\mu^2_\text{h}$ and $\mu^2_\text{s}$ can be fixed by the stationary conditions for $(v,w)$ at a given scale $\mu=\Lambda$,
\begin{equation}\label{muhmus}
\begin{split}
 &\mu^2_\text{h}= \lambda_\text{h} v^2+ \lambda_\text{hs} w^2+ \kappa_\text{hs} w\\
& \mu^2_\text{s}= \lambda_\text{s} w^2+ \lambda_\text{hs} v^2+ \kappa_\text{s} w+\frac{\kappa_\text{hs} v^2}{2w}
 \end{split}
\end{equation}
where the Higgs VEV is fixed at $v=246$ GeV at the EW scale, and for the scalar VEV we take benchmark values $w= 100$ GeV and $w=1, 10, 100$ TeV at the EW scale. Note that we choose the free parameters of the model to be $\lambda_\text{h}, \lambda_\text{s}, \lambda_\text{hs},  \kappa_\text{h}, \kappa_\text{hs}$; we do not use a mixing angle as a new free parameter, instead we directly deal with the various $(v,w)$ inputs from scratch and investigate the properties of the model based on the chosen vacuum.  
\begin{table}
 \begin{center}
 \small   \begin{tabular}{ | l | l | l  |l| l| l| l| p{2cm} |}
    \hline
    $w$  & $100$ GeV&$1$ TeV&$10$ TeV& $100 $ TeV\\ \hline 
 $\lambda_\text{h}$    & $(0.02,0.24)$ & $(0.002,0.222)$ & $(0.00054,0.209)$  & $(8.5\times 10^{-5},0.096)$  \\ \hline
   $\lambda_\text{s}$ & $(0.002,0.97)$ & $(2.26\times 10^{-6} , 0.014)$& $(2.54\times 10^{-6} , 0.00014)$ & $(3.10 , 11.45)\times 10^{-7}$  \\ \hline
     $\lambda_\text{hs}$ &$(-0.098,0.098)$  &$(-0.019,0.02)$ & $(-0.0021,0.0023)$ &$(-7.73,8.90)\times 10^{-6}$  \\ \hline
     $\kappa_\text{s}$  & $(-0.98,0.99)$ &$(-1,0.97)$ & $(-0.96,0.99)$ &$(-0.25,-0.05)$\\ \hline
      $\kappa_\text{hs}$  & $ (-1,1)$& $(-0.95,0.99)$& $(-0.96,0.98)$ & $(-0.71,0.95)$\\ \hline
      $m_s$ & $48$-$119$ GeV & $46$-$118$ GeV &$4$-$123$ GeV& $8$-$109$ GeV \\ \hline
      \end{tabular}
\end{center}
        \caption{The allowed region for the couplings $\lambda_\text{h},\lambda_\text{s},\lambda_\text{hs}$, $\kappa_\text{s},\kappa_\text{hs}$ and $m_s$ for different singlet scalar VEV benchmarks $w=0.1, 1, 10, 100$ TeV, respecting the absolute vacuum stability and positivity condition for the vacuum $(v,w)$ at the electroweak scale with the assumptions $m_H=125$ GeV, $v_H=246$ GeV and $m_s<m_H$.}
        \label{PTtable2}
        \end{table} 
        
The Hessian matrix at a given scale $\mu=\Lambda$ in the scale-dependent vacuum $(v,w)$ is given by, 
\begin{equation}\label{hessian}
\mathcal{H}(v,w;\Lambda)=\left(
 \begin{matrix}
  3 \lambda_\text{h}  v^2 +  \lambda_\text{hs} w^2 + \kappa_\text{hs} w- \mu^2_\text{h} & 2  \lambda_\text{hs}v w + \kappa_\text{hs} v\\
 2 \lambda_\text{hs} v w +\kappa_\text{hs} v &  3  \lambda_\text{s} w^2+ \lambda_\text{hs}v^2 +  2\kappa_\text{s} w- \mu^2_\text{s}\\
 \end{matrix}\right)
\end{equation}
where we have dropped in the matrix the scale-dependence of the couplings and the VEVs. Taking into account the stationary condition on the vacuum $(v,w)$ in Eq. (\ref{muhmus}) the mass eigenvalues are,
\begin{equation}\label{mas}
\begin{split}
 &m_{\pm}^2= v^2  \lambda_\text{h} + w^2  \lambda_\text{s} - \frac{v^2}{4w}  \kappa_\text{hs} +\frac{w}{2}  \kappa_\text{s}\pm \frac{v}{2}\times \\
 &\sqrt{\frac{v^2}{w^2}\left(\kappa_\text{hs} +4w \lambda_\text{h}\right) +\frac{w^2}{v^2}\left(\kappa_\text{s} 
+2w \lambda_\text{s}\right)+2 \kappa_\text{hs} w\left(\lambda_\text{s}-8\lambda_\text{hs}\right)+4w\left(\kappa_\text{s} \lambda_\text{h} -4w\lambda_\text{hs}^2+2w \lambda_\text{h}  \lambda_\text{s}  \right)-4\kappa_\text{hs}^2}\\
 \end{split}
\end{equation}
% % % % 
 
% % % % 
where $m_-$ and $m_+$ can be either the Higgs or the scalar mass. In section \ref{numeric}, we will consider both possibilities $m_s<m_H$ and $m_s>m_H$ at the EW scale. Although we set the initial inputs of the free parameters such that $(v,w)$ is the absolute minimum at the EW scale, but running the couplings in higher energy scales, the vacuum $(0,0)$ or $(0,w)$ may become coexistent minima of the theory even deeper than the $(v,w)$, which results in the instability of the vacuum at higher scales. 
To keep the $(v,w)$ to be the absolute minimum at higher scales, we need to know the mass spectrum of other possible vacua in higher scales. The mass matrix for the vacuum $(0,0)$ at a given scale $\Lambda$ read
\begin{equation}\label{mas00}
 \mathcal{M}(0,0;\Lambda)=\begin{pmatrix}
 - \mu^2_\text{h}& 0\\
  0 & -\mu^2_\text{s}\\
 \end{pmatrix}
\end{equation}

and the mass matrix for the vacuum $(0,w)$ is
\begin{equation}\label{mas0w}
 \mathcal{M}(0,w;\Lambda)=\begin{pmatrix}
 w^2 \lambda_\text{hs} + w \kappa_\text{hs}- \mu^2_\text{h}& 0\\
  0 &2 w^2 \lambda_\text{s}+ w \kappa_\text{s}\\
 \end{pmatrix}
\end{equation}
% % % % 
% % % % 
in which the stationary condition
\begin{equation}
 \mu^2_\text{s}= w^2 \lambda_\text{s}+ w \kappa_\text{s}
\end{equation}
 for $(0,w)$ has been used. The initial values for the parameters $\mu^2_\text{h}$ and $\mu^2_\text{s}$ in Eqs. (\ref{mas00}) and (\ref{mas0w}) is given by Eq. (\ref{muhmus}). In order for $(v,w)$ to stay the global minimum up to a desired scale at least one of the mass eigenvalues in Eqs. (\ref{mas00}) and (\ref{mas0w}) must be negative; in this way $(v,w)$ would be an absolute minimum. Already at $\Lambda= \mathcal{O}(m_t)\sim 173$ GeV, from Eq. (\ref{mas00}) we must require  $\mu^2_\text{h}>0$ or $\mu^2_\text{s}>0$ and from Eq. (\ref{mas0w}), $\mu^2_\text{h}>w^2 \lambda_\text{hs} + w \kappa_\text{hs}$ or $2 w^2 \lambda_\text{s}+ w \kappa_\text{s}<0$ in which both parameters $\mu^2_\text{s}$ and $\mu^2_\text{s}$ are fixed at $\Lambda=173$ GeV from Eq. (\ref{muhmus}).

\section{Renormalizarion Group Equations}\label{rge}
\begin{figure}
\begin{multicols}{2}
    \includegraphics[width=\linewidth]{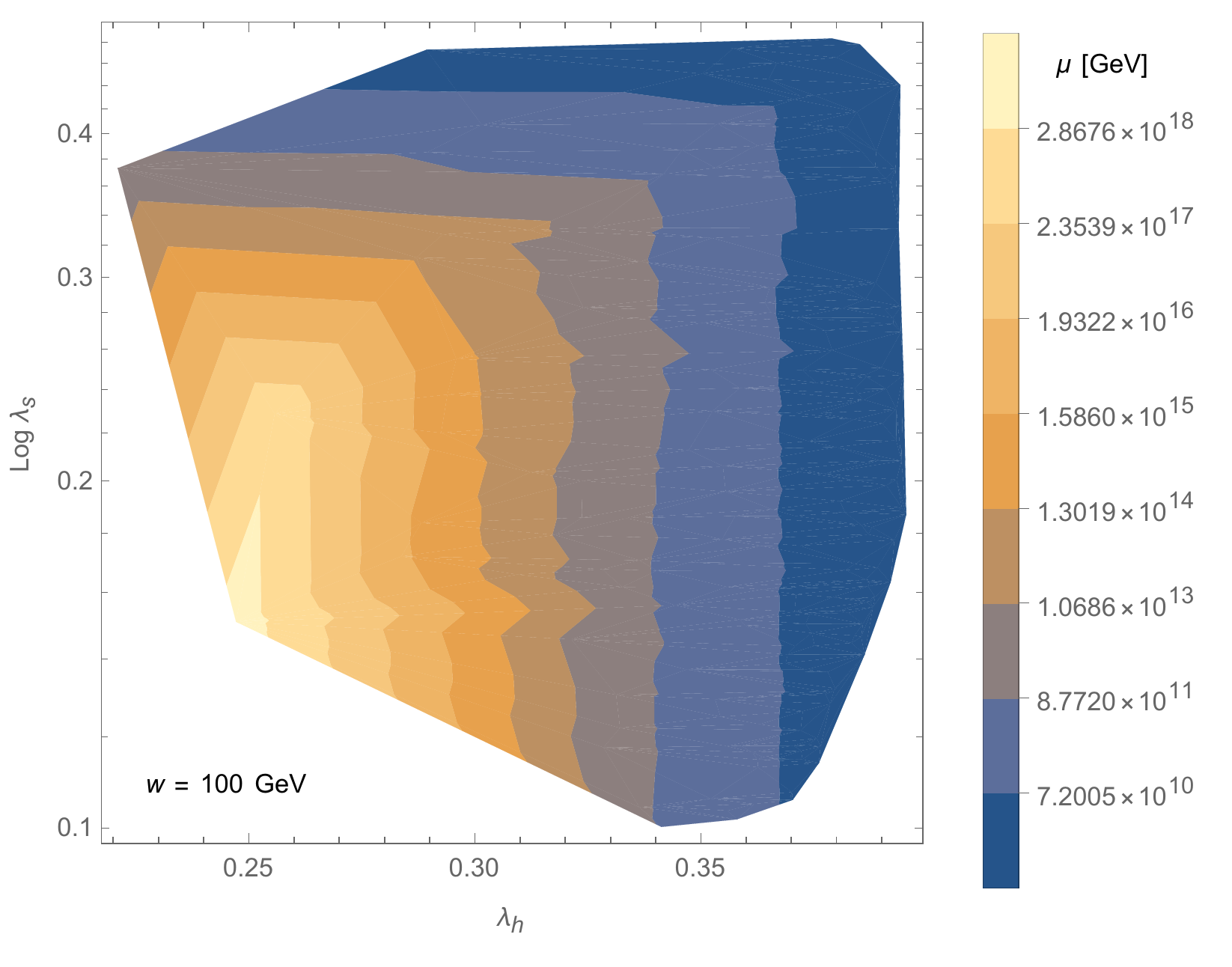}\par 
    \includegraphics[width=\linewidth]{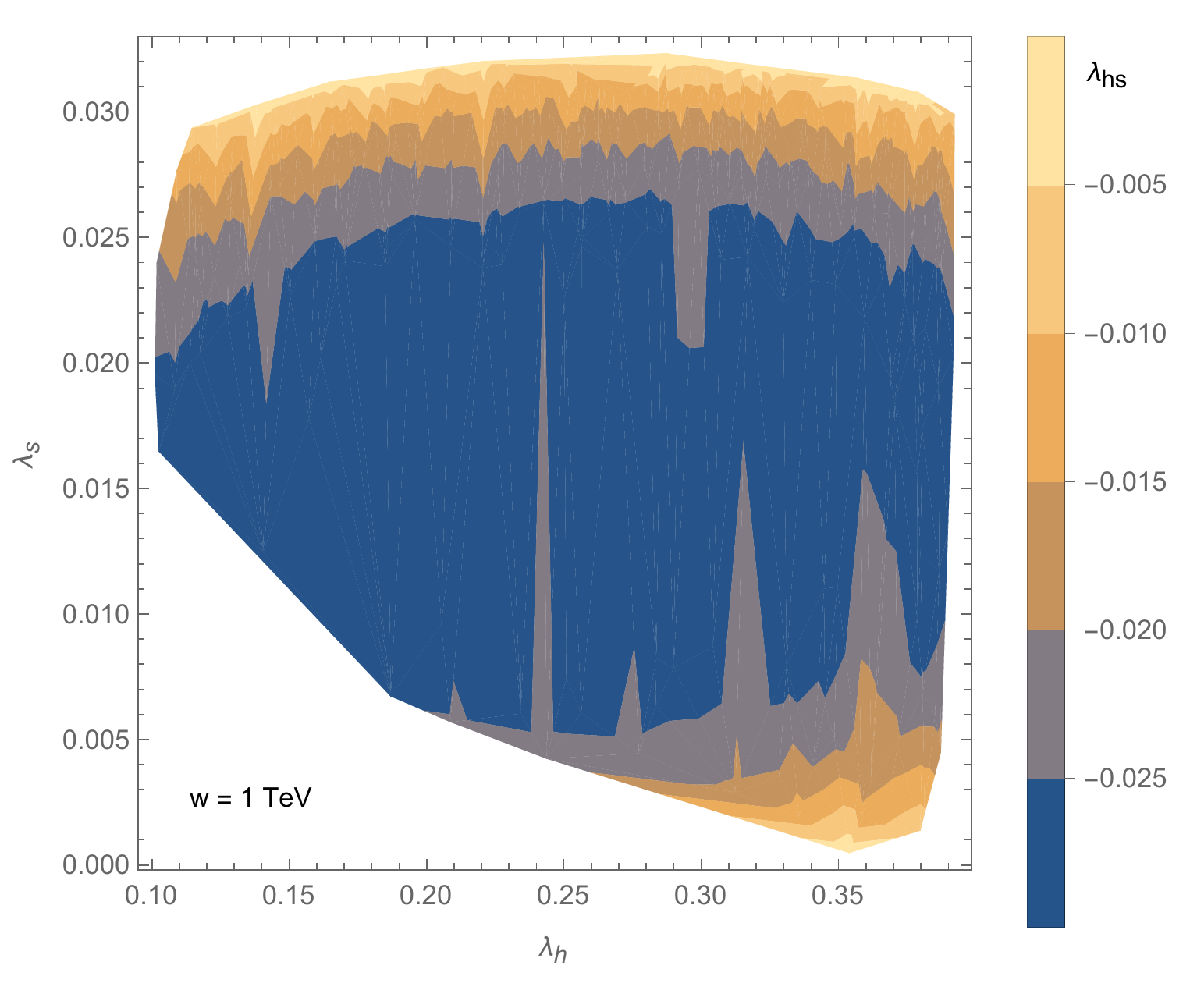}\par 
    \end{multicols}
\begin{multicols}{2}
    \includegraphics[width=\linewidth]{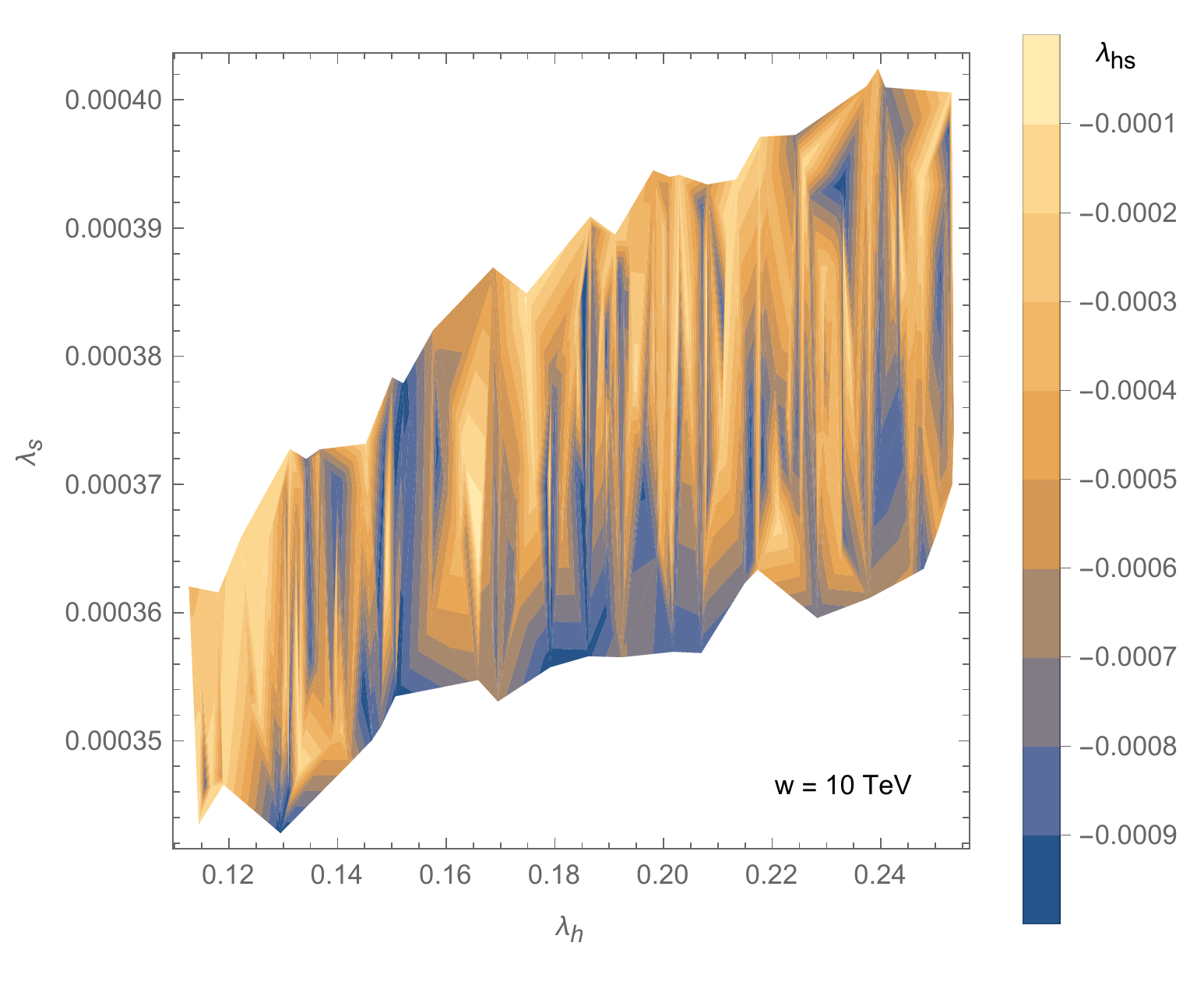}\par
    \includegraphics[width=\linewidth]{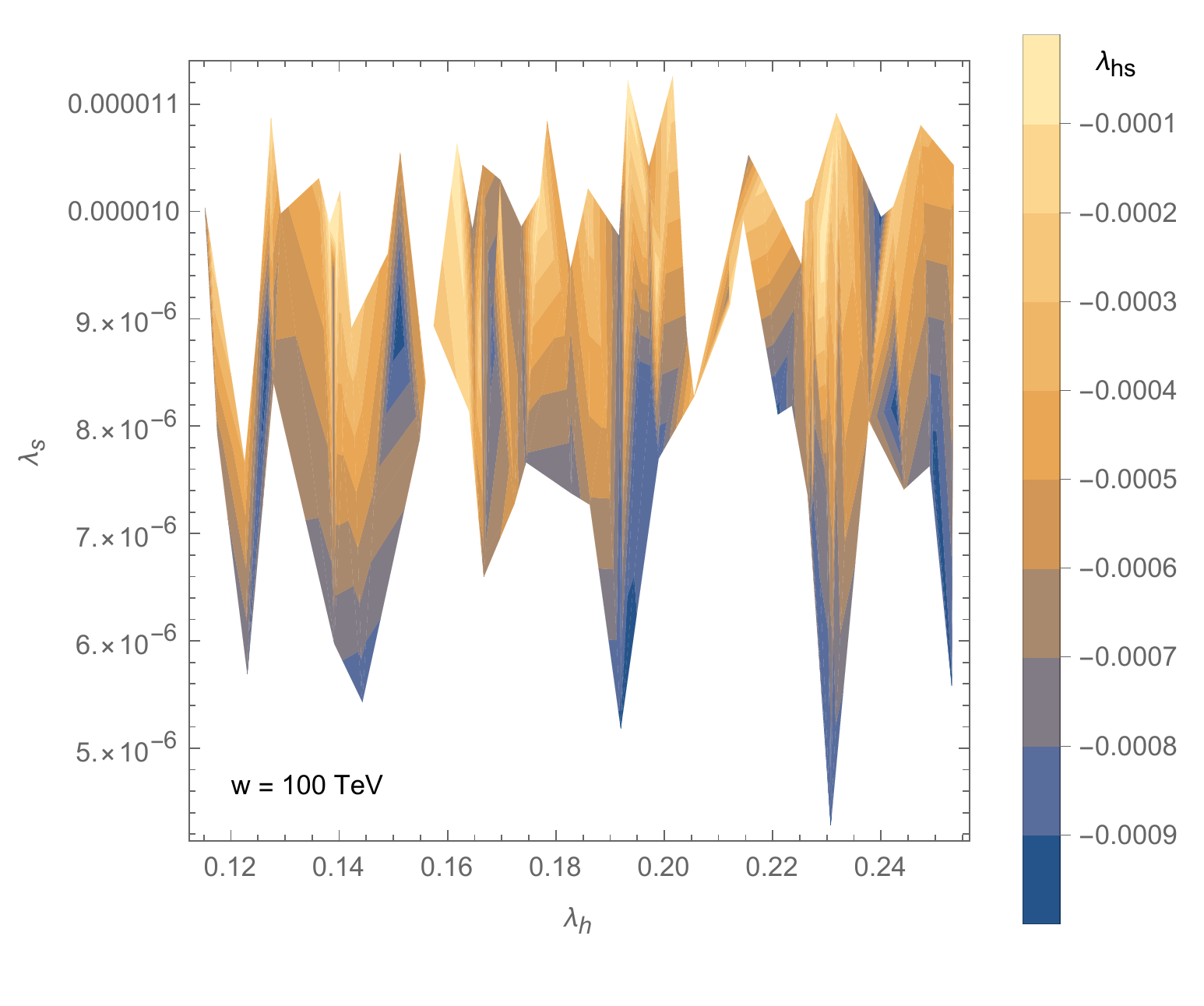}\par
\end{multicols} 
\caption{The viable region of input values for the couplings $\lambda_\text{h}$, $\lambda_\text{s}$ and $\lambda_\text{hs}$ at the 
electroweak scale for different singlet scalar VEV benchmarks $w=0.1, 1, 10, 100$ TeV, respecting the absolute vacuum stability for the vacuum $(v,w)$, the positinity condition and the perturbativity up to the Planck scale, with the assumptions $m_H=125$ GeV, $v=246$ GeV and  $m_s>m_H$.}\label{regms>mh}
\end{figure}

The evolution of the couplings, fields and mass parameters in the model with scale $\mu$ is given by the renormalization group equations (RGE). We extract the RGEs for the model given in Eq. (\ref{sinpot}) up to one-loop using the Mathematica package {\fontfamily{cmtt}\selectfont SARAH} \cite{Staub:2015kfa}. We take into account only the top quark Yukawa coupling and ignore the couplings for the light quarks. The $\beta$-functions for the gauge couplings and the Yukawa coupling are, 
\begin{equation}\label{gyrge}
\begin{split}
 16\pi^2 &\beta_{g_i}=b_i g_i^3 \\
 16\pi^2 &\beta_{y_t}=\left(-\frac{17}{20} g^2_1-\frac{9}{4} g_2^2  -8g_3^2\right) y_t+ \frac{3}{2} y_t^3\\
\end{split}
\end{equation}
% % 
% % 
where $b_1=41/10,~b_2=-19/6,~b_3=-7$. The $\beta$-functions involving the Higgs and the singlet scalar  couplings are given by,
\begin{equation}\label{cplrge}
\begin{split}
& 16\pi^2  \beta_{\lambda_\text{h}}=\left( \frac{27}{200}g_1^4+\frac{9}{20} g_1^2 g_2^2 +\frac{9}{8} g_2^4\right) +\left(-\frac{9}{5} g_1^2  - 9 g_2^2 +12 y_t^2\right) \lambda_\text{h} +\frac{1}{2} \lambda_\text{hs}^2+24 \lambda_\text{h}^2  -6 y_t^4\\
 &16\pi^2  \beta_{\lambda_\text{s}}=2\lambda_\text{hs}^2 +18 \lambda^2_\text{s}\\
 &16\pi^2  \beta_{\lambda_\text{hs}}= \left(-\frac{9}{10} g^2_1-\frac{9}{2} g_2^2  +6 \lambda_\text{s} +12\lambda_\text{h} + 6 y_t^2 \right)\lambda_\text{hs} +4\lambda_\text{hs}^2\\
  & 16\pi^2  \beta_{\kappa_\text{s}} = 6 \kappa_\text{hs} \lambda_\text{hs} +18 \lambda_\text{h} \kappa_\text{s} \\
 & 16\pi^2  \beta_{\kappa_\text{hs}}= \left( -\frac{9}{10} g_1^2  -\frac{9}{2} g_2^2  + 4  \lambda_\text{hs}+12 \lambda_\text{h} +6  y_t^2 \right)\kappa_\text{hs}  + 2\kappa_\text{s}  \lambda_\text{hs} \\
\end{split}
\end{equation}
% % % 
\begin{figure}
\begin{multicols}{2}
    \includegraphics[width=\linewidth]{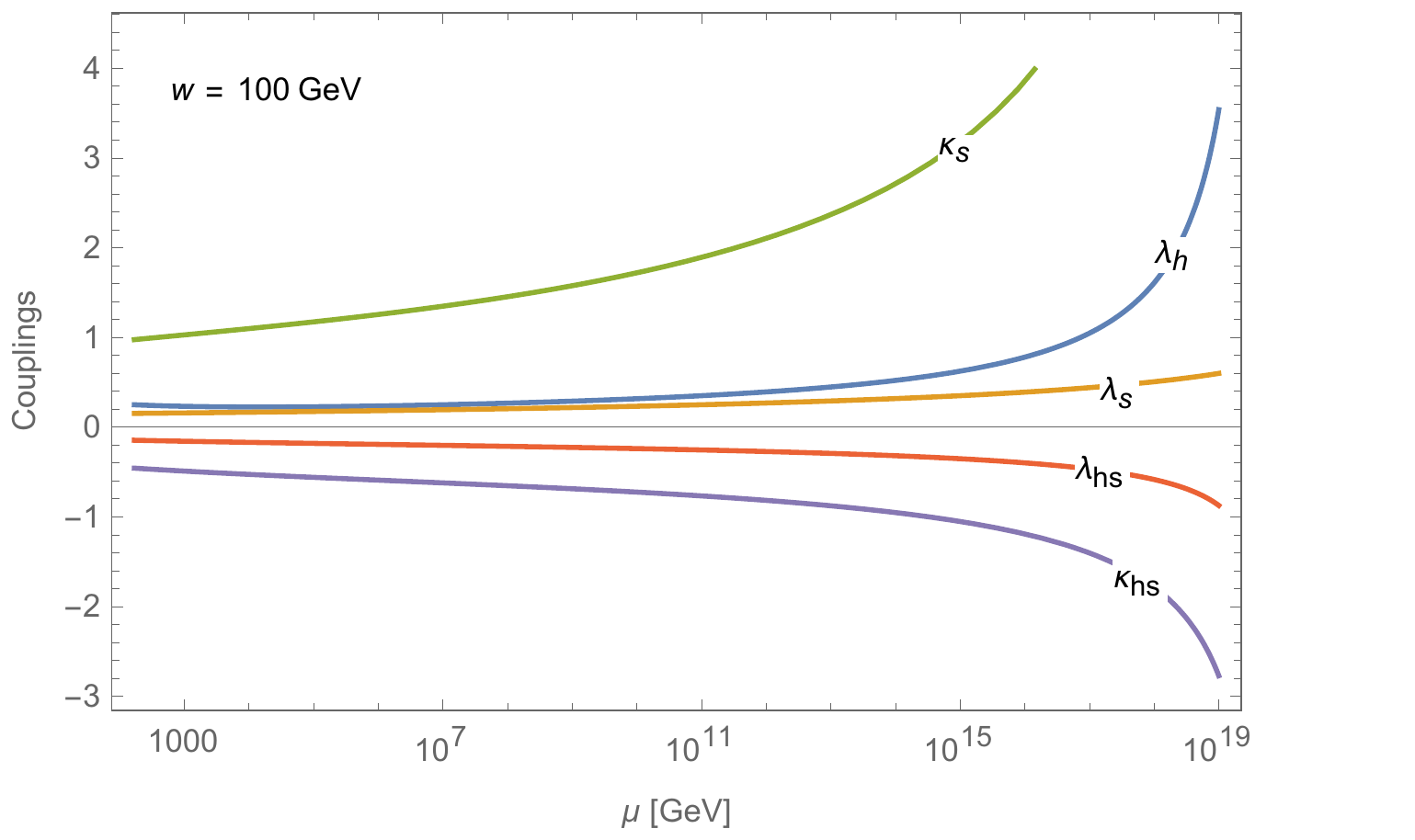}\par 
    \includegraphics[width=\linewidth]{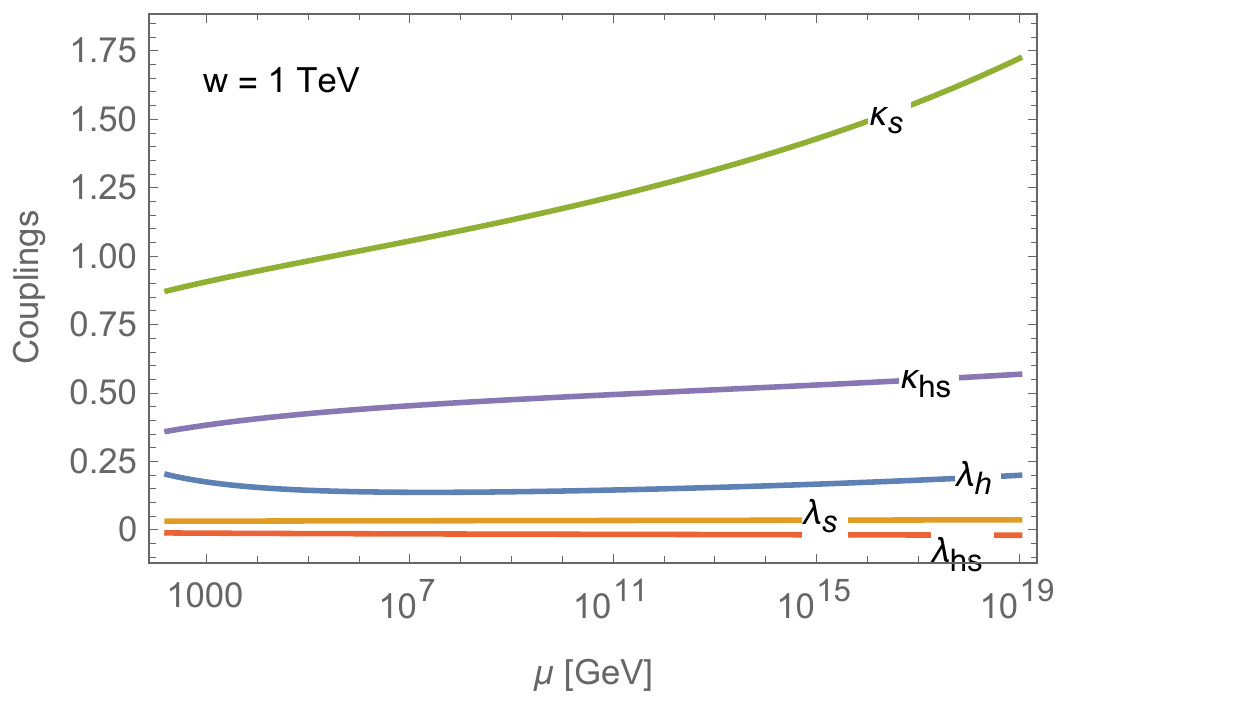}\par 
    \end{multicols}
\begin{multicols}{2}
    \includegraphics[width=\linewidth]{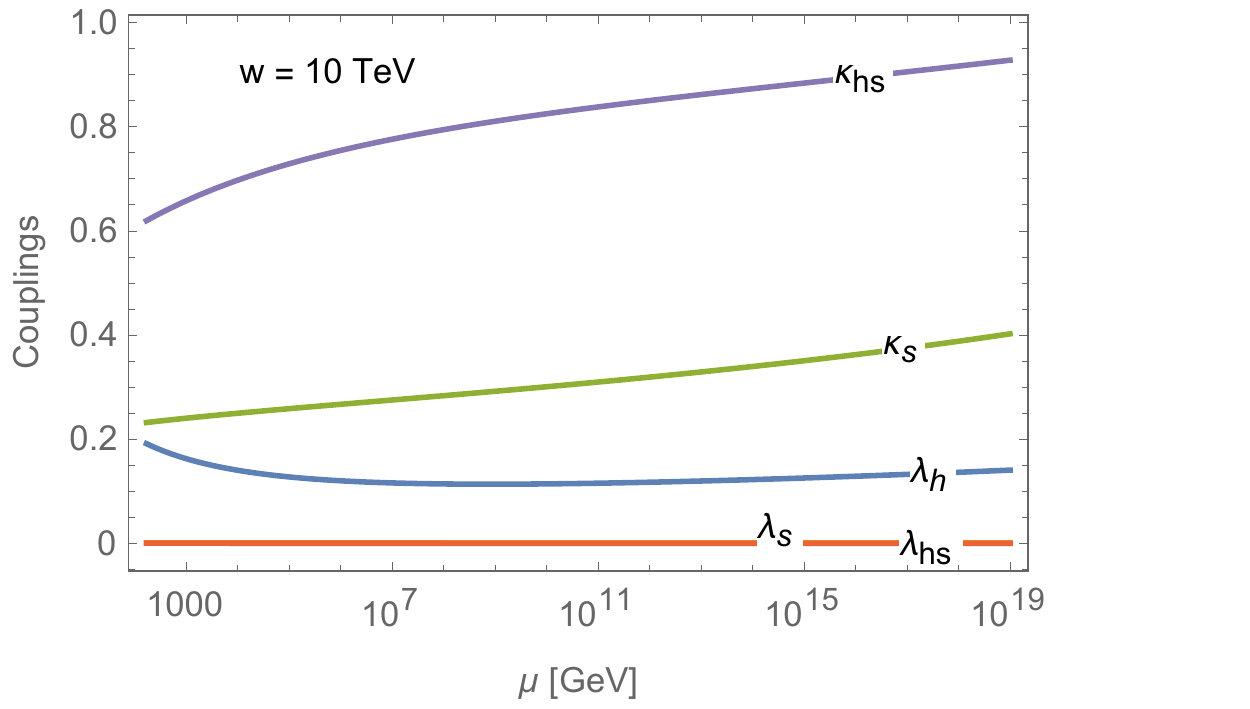}\par
    \includegraphics[width=\linewidth]{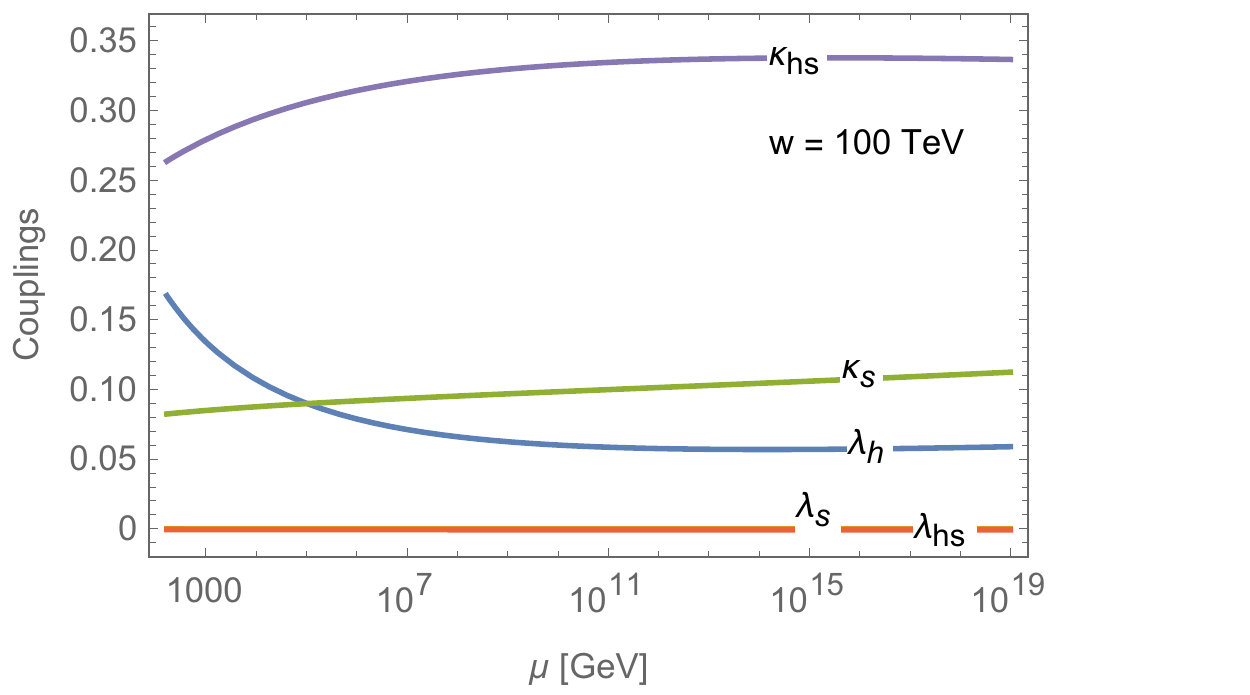}\par
\end{multicols}
\caption{The plots show the running of the couplings $\lambda_\text{h},\lambda_\text{s},\lambda_\text{hs},\kappa_\text{s},\kappa_\text{hs}$ up to the Planck scale for different singlet scalar VEV benchmarks  $w=0.1, 1, 10, 100$ TeV and with the constraints $m_H=125$ GeV, $v=246$ GeV and $m_s>m_h$. The initial inputs for the couplings are chosen such that the vacuum $(v,w)$ remains the absolute minimum respecting the positivity condition and the perturbativity up to the Plack scale.
}\label{runms>mh}
\end{figure}

% % % 
and the $\gamma$-functions for the VEVs and the mass parameters read, 
\begin{equation}\label{mpm}
\begin{split}
 &16\pi^2  \gamma_v=\left( \frac{9}{20} g_1^2+\frac{9}{4}g_2^2-3y_t^2\right) v  \\
 &16\pi^2  \gamma_w=0 \\
 &16\pi^2  \gamma_{\mu^2_\text{h}}= \left(-\frac{9}{10} g_1^2 -\frac{9}{2}g_2^2 + 12 \lambda_\text{h} +6 y_t^2 \right) \mu^2_\text{h} -2 \kappa_\text{hs}^2 +\lambda_\text{hs} \mu^2_\text{s}\\
 &16\pi^2  \gamma_{\mu^2_\text{s}}=4 \mu^2_\text{h} \lambda_\text{hs}+6\lambda_\text{s}\mu^2_\text{s}-4\kappa_\text{hs}^2-4\kappa_\text{s}^2
\end{split}
\end{equation}
where the $\beta$-functions for a coupling $X$, and the $\gamma$-functions (anomalous dimensions) for VEV or mass parameter $Y$, are defined as,
\begin{equation}
 \beta_X= \mu \frac{dX}{d\mu}\hspace{3cm}\gamma_Y=-\frac{\mu}{Y}\frac{dY}{d\mu} \,.
\end{equation}

Solving the RGEs in Eqs. (\ref{gyrge}), (\ref{cplrge}) and (\ref{mpm}), the evolution of the couplings and the VEVs will be known, so we can check the status of the stability, the positivity and the perturbativity all together at any desired scale. In the next section we will discuss the RGE solutions and the allowed parameters inputs that we can use at the EW scale. 
 
\begin{figure}
\begin{multicols}{2}
    \includegraphics[width=\linewidth]{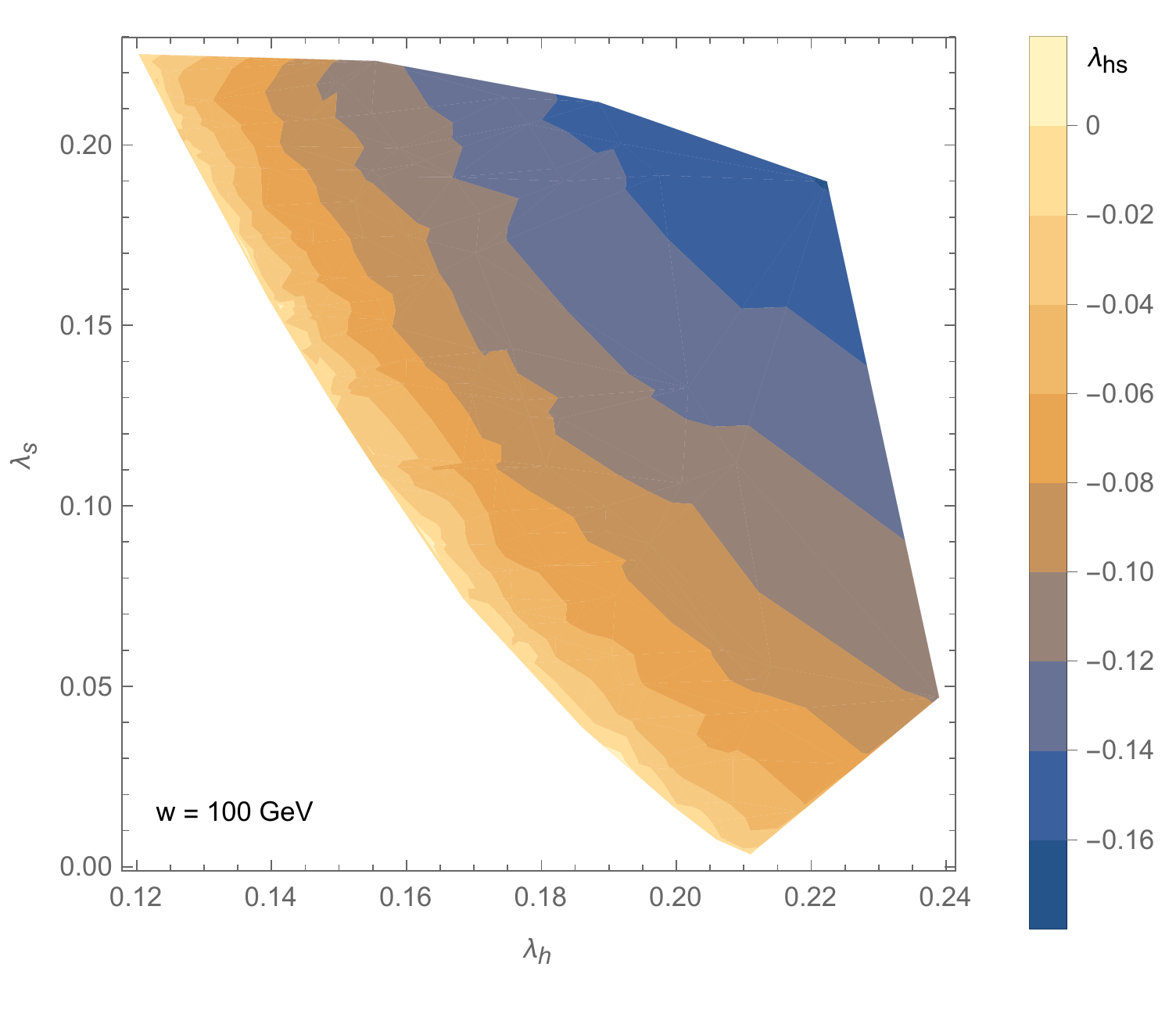}\par 
    \includegraphics[width=\linewidth]{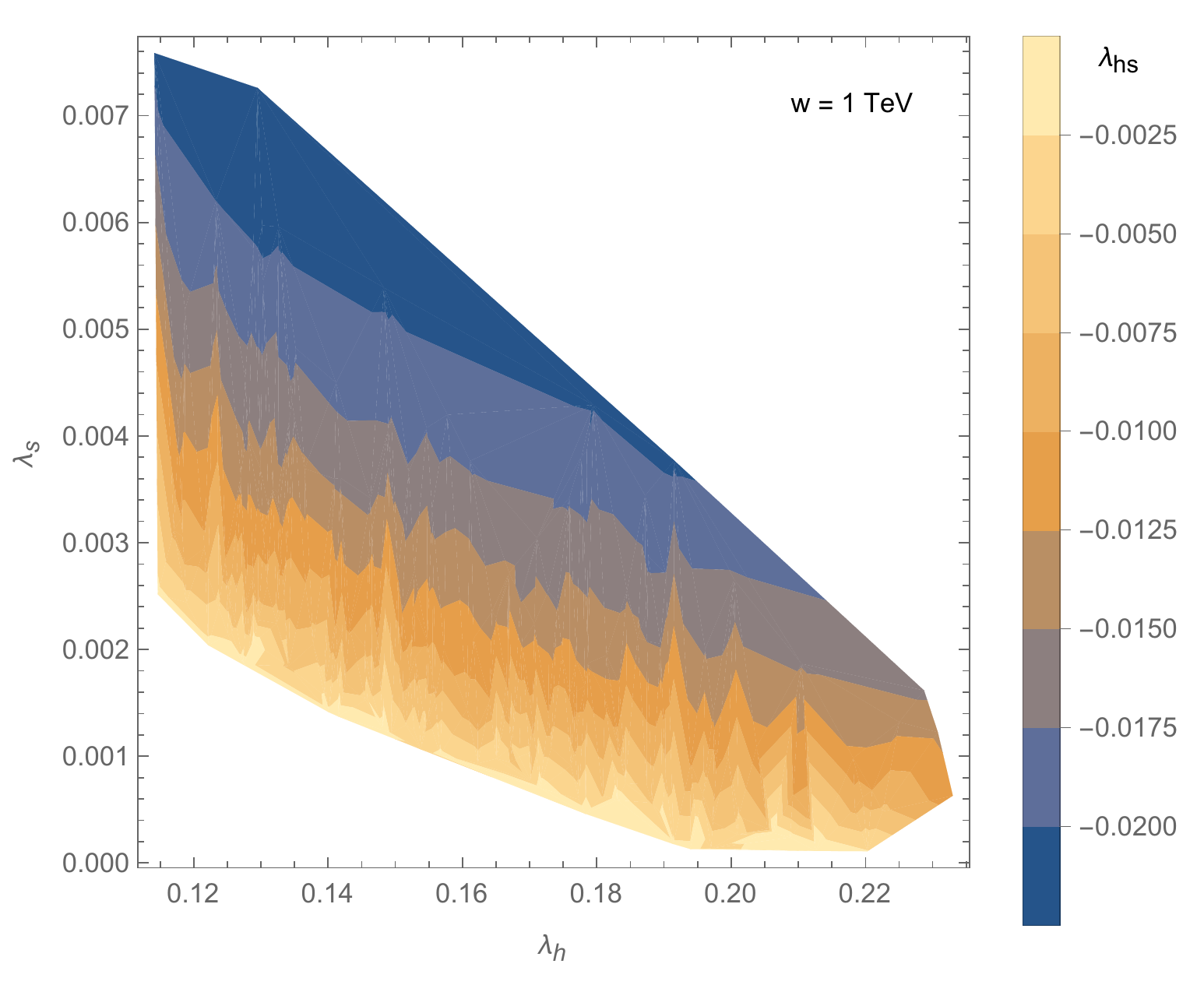}\par 
    \end{multicols}
\begin{multicols}{2}
    \includegraphics[width=\linewidth]{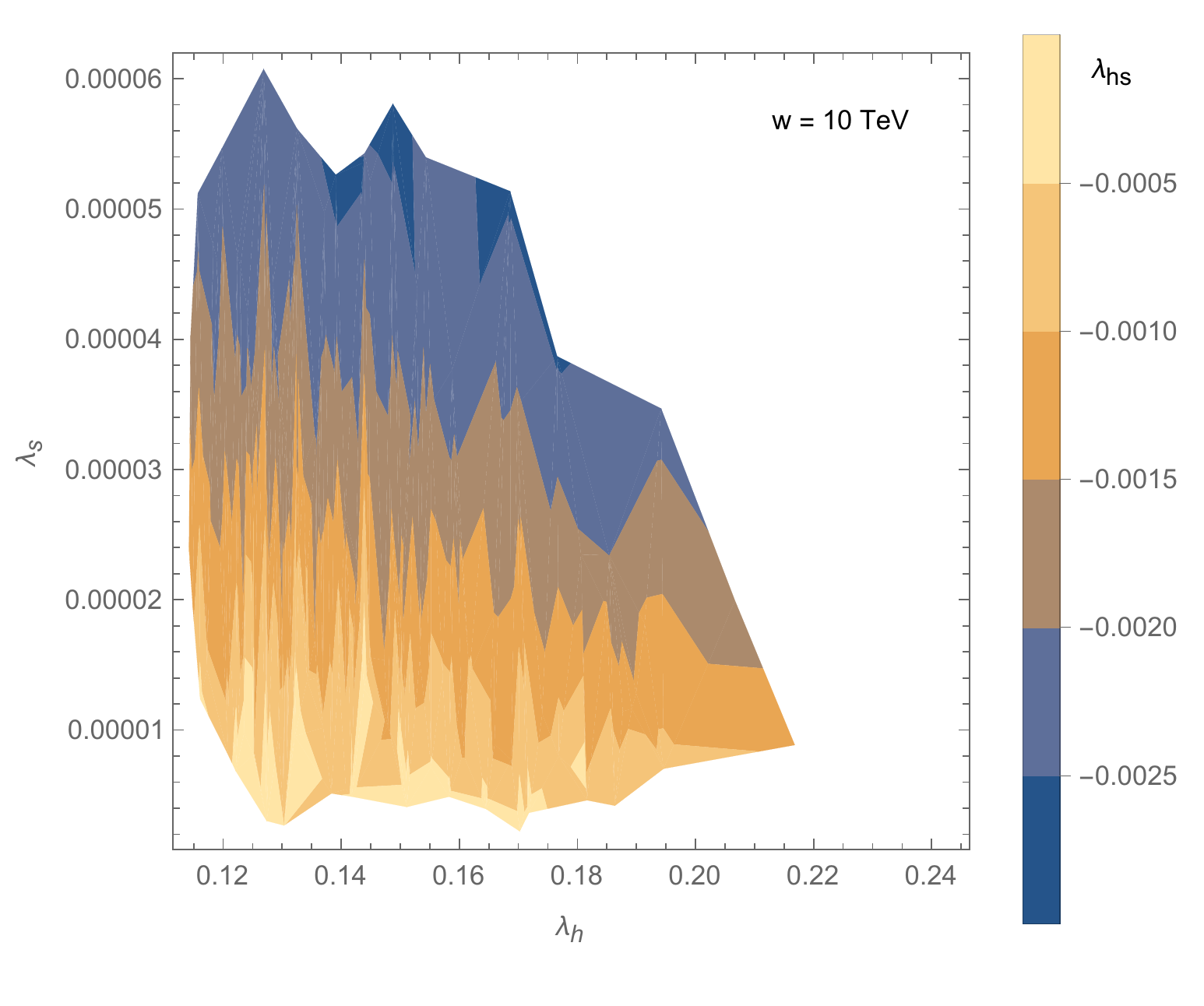}\par
    \includegraphics[width=\linewidth]{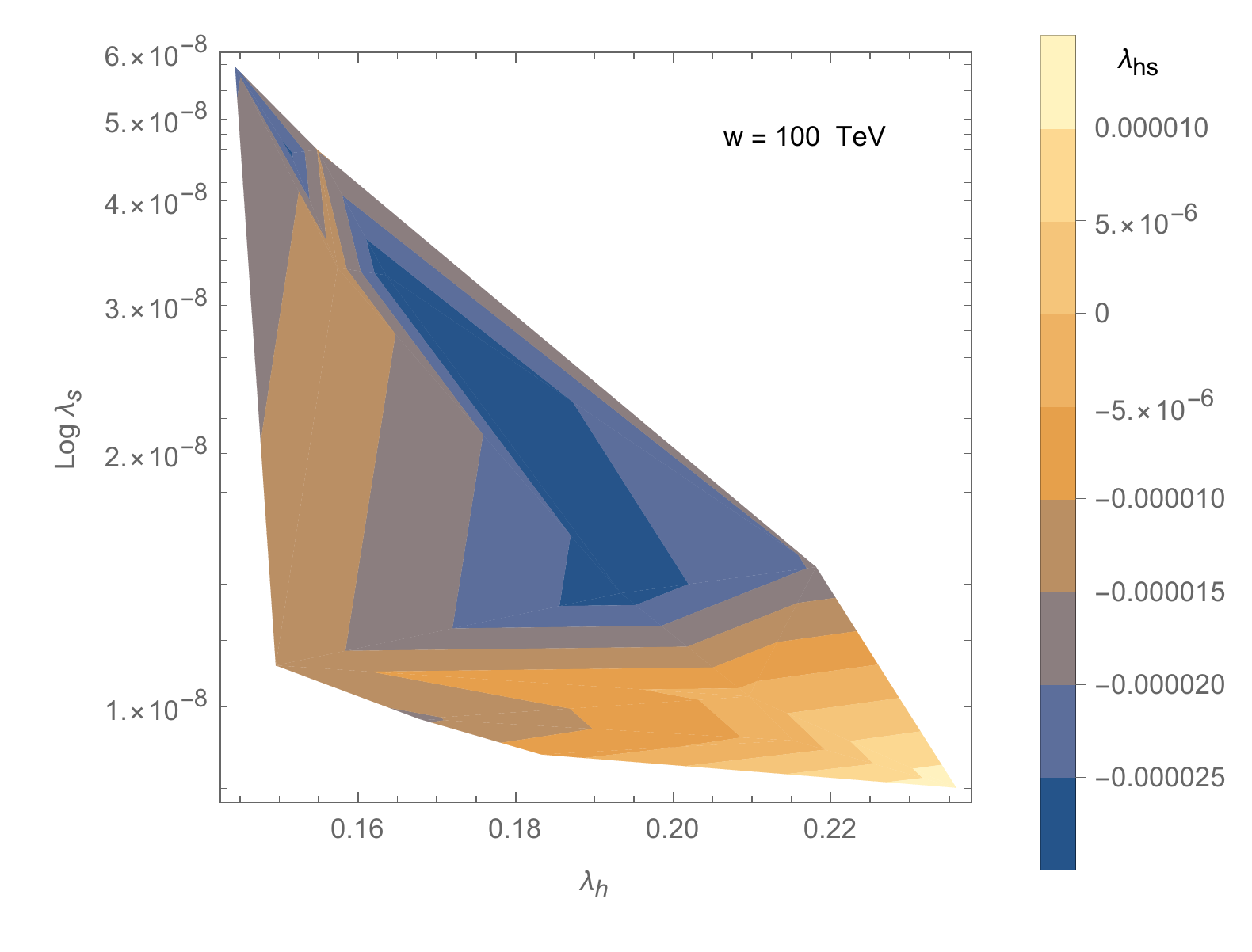}\par
\end{multicols} 
\caption{The viable region of input values for the couplings $\lambda_\text{h}$, $\lambda_\text{s}$ and $\lambda_\text{hs}$ at the 
electroweak scale for different singlet scalar VEV benchmarks $w=0.1, 1, 10, 100$ TeV, respecting the absolute vacuum stability for the vacuum $(v,w)$, the positinity condition and the perturbativity up to the Planck scale, with the assumptions $m_H=125$ GeV, $v=246$ GeV and  $m_s<m_H$.}\label{regms<mh}
\end{figure}

\section{Vacuum Stability, Positivity and Perturbativity}\label{numeric}

In this section we numerically solve the RG equations presented in section \ref{rge}. We always require that the vacuum $(v,w)$ defined in the EW scale $\mathcal{O}(m_t)\sim 173$ GeV, remains the absolute global minimum for higher scales up to the Planck scale, hence pushing the stability up to the Planck scale. Furthermore, we impose the positivity condition in Eq. (\ref{pos}) (which is an scale-dependent condition evolving with the couplings) to hold from the EW scale up to the Planck scale. We also discard the input values for the parameters which lead to a Landau pole in the scales lower than the Planck scale. 

\begin{figure}
\begin{multicols}{2}
    \includegraphics[width=\linewidth]{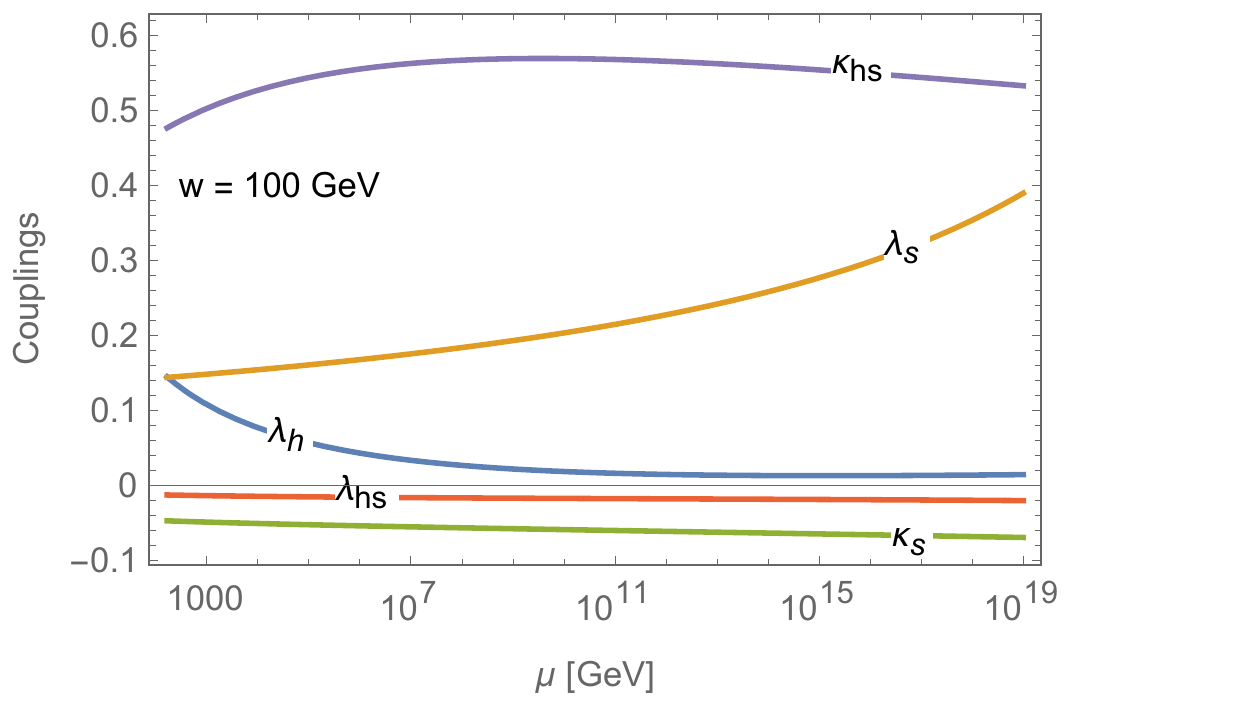}\par 
    \includegraphics[width=\linewidth]{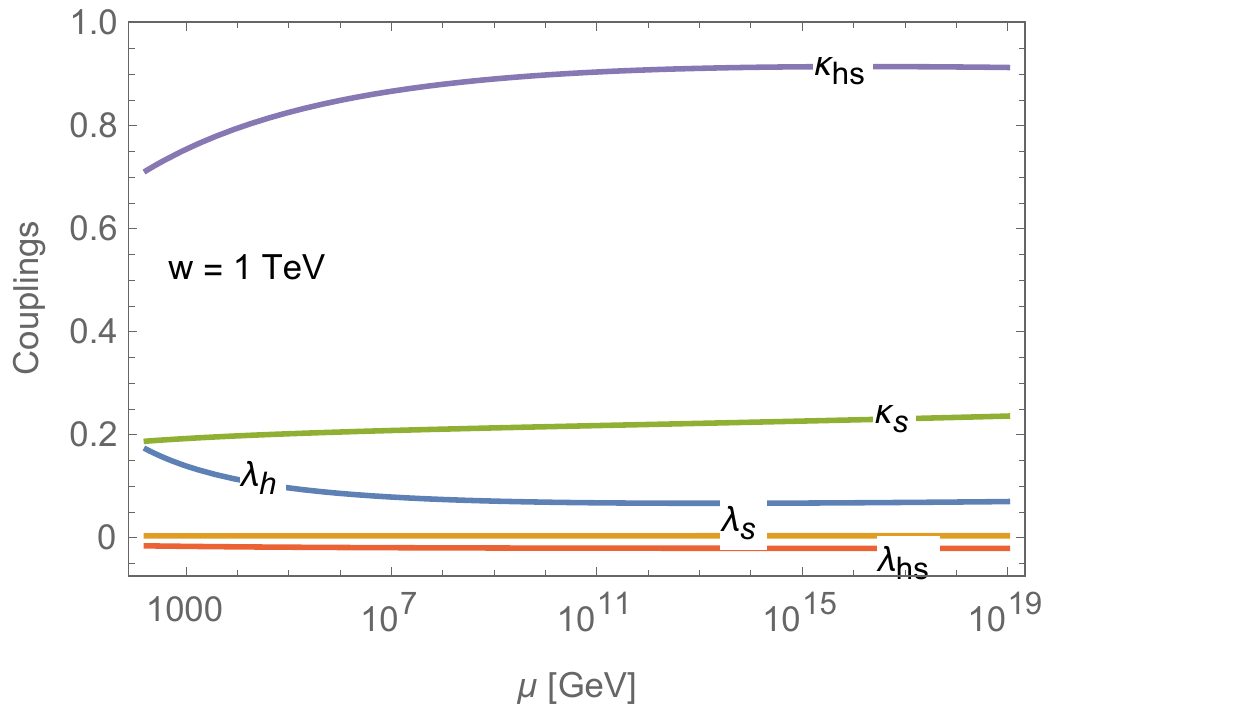}\par 
    \end{multicols}
\begin{multicols}{2}
    \includegraphics[width=\linewidth]{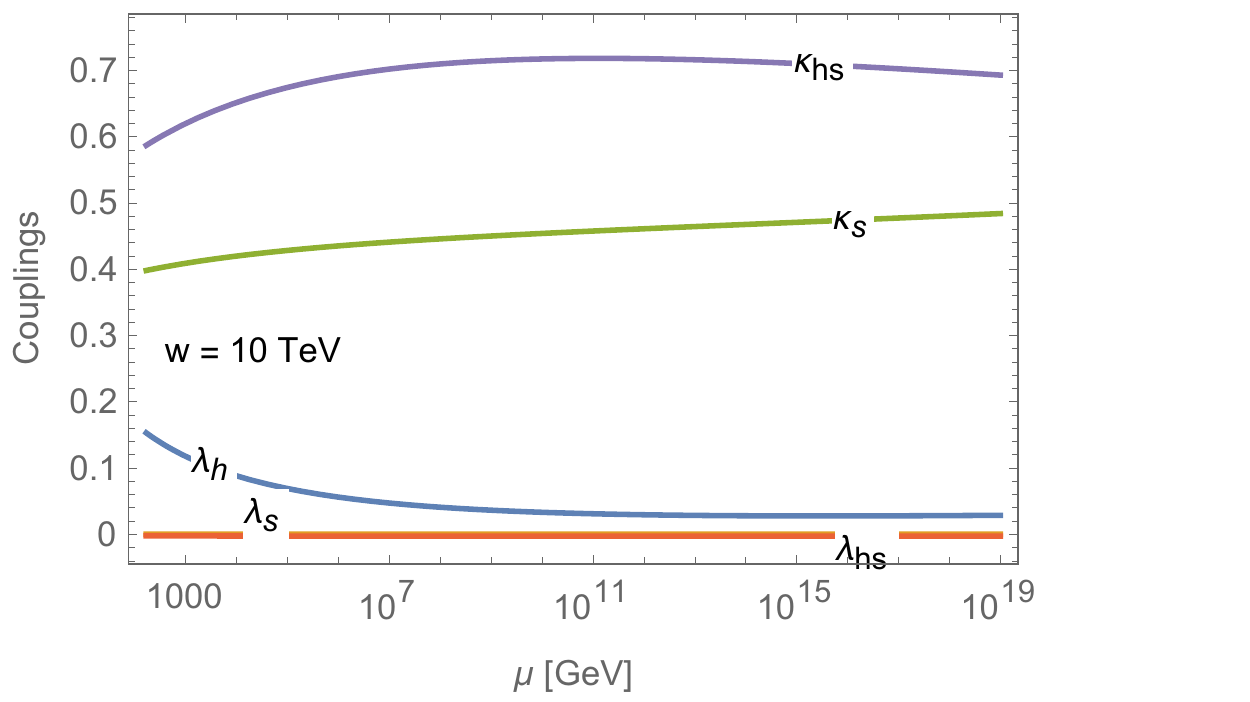}\par
    \includegraphics[width=\linewidth]{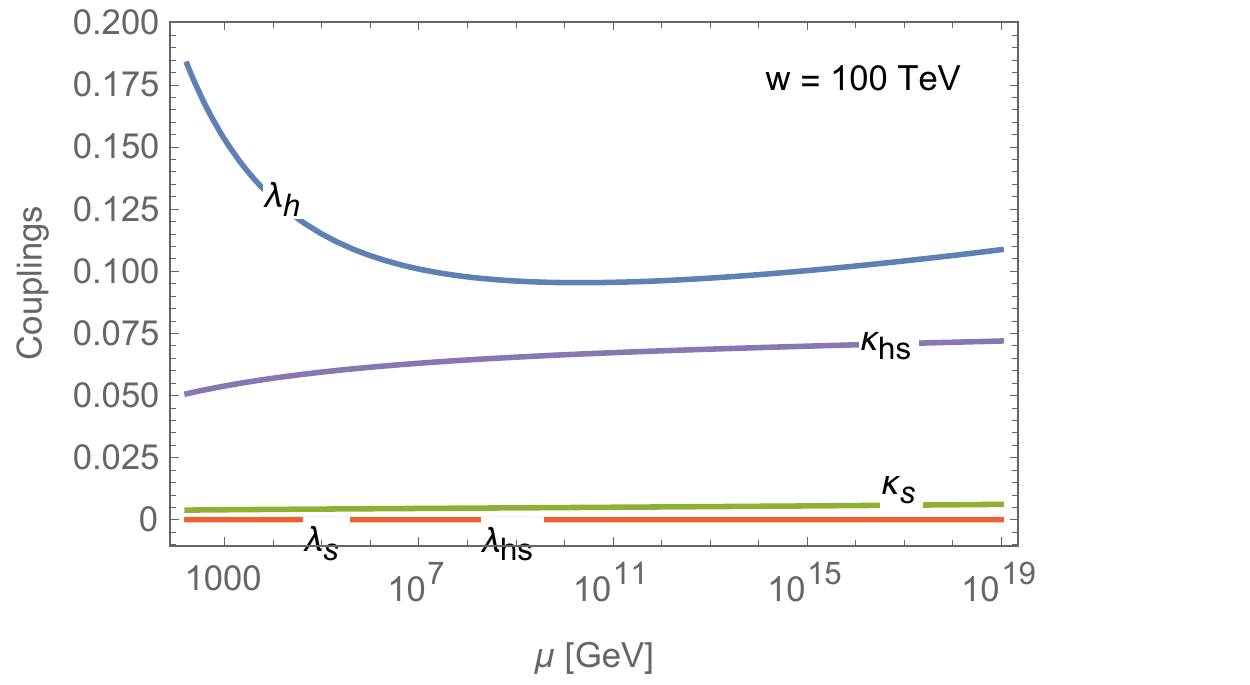}\par 
\end{multicols}
\caption{The plots show the running of the couplings $\lambda_\text{h},\lambda_\text{s},\lambda_\text{hs},\kappa_\text{s},\kappa_\text{hs}$ up to the Planck scale for different singlet scalar VEV benchmarks  $w=0.1, 1, 10, 100$ TeV and with the constraints $m_H=125$ GeV, $v=246$ GeV and $m_s<m_h$. The initial inputs for the couplings are chosen such that the vacuum $(v,w)$ remains the absolute minimum respecting the positivity condition and the perturbativity up to the Plack scale.}\label{runms<mh}
\end{figure}

From the LHC experiments, the Higgs mass and the Higgs VEV are known: $v=246$ GeV and $m_H=125$ GeV. We will also fix the singlet scalar VEV by different mass scale benchmarks $w=0.1, 1, 2 , 10 , 100$ TeV. In Eq. (\ref{mas}) one of the mass eigenvalues is attributed to the Higgs mass. We investigate both cases $m_+\equiv m_s> m_H\equiv m_-$ and $m_-\equiv m_s<m_H\equiv m_+$. Among the parameters of the model as seen in section \ref{model}, $\mu^2_\text{h}$ and $\mu^2_\text{s}$ are omitted by two stationary conditions for $(v,w)$. 
Doing so we are left with the free independent parameters being $\lambda_\text{h}$, $\lambda_\text{s}$, $\lambda_\text{hs}$, $\kappa_\text{s}$ and $\kappa_\text{hs}$. 
The input values for the free parameters at $\mathcal{O}(m_t)$ scale should be chosen such that the vacuum $(v,w)$ be the absolute global minimum. Any set of input for the free parameters will give an input for $\mu^2_\text{h}$ and $\mu^2_\text{s}$ from Eq. (\ref{muhmus}). Using these values in Eqs. (\ref{mas00}) and (\ref{mas0w}) at least one of the mass eigenvalues in $\mathcal{M}(0,0)$ and in $\mathcal{M}(0,w)$ must be negative. Moreover, 
at the EW scale the input values chosen for the free parameters should be bounded by the positivity condition in Eq. (\ref{pos}), the positivity of the radicand in the mass expressions in Eq. (\ref{mpm}), and the positivity of the mass eigenvalue $m_-$ in Eq. (\ref{mas}). Also depending on taking $m_H\equiv m_+\sim 125$ GeV or  $m_H\equiv m_-\sim 125$ GeV, the free parameters are bounded differently. 
The initial values for the set of parameters satisfying the aforementioned constrained are presented in Table \ref{PTtable1} for $m_s>m_H$, and in Table \ref{PTtable2} for $m_s<m_H$ with the benchmarks $w=0.1, 1, 10, 100$ TeV. In both tables the range of the allowed singlet scale mass is shown. All the parameters $\lambda_\text{h}$, $\lambda_\text{s}$, $\lambda_\text{hs}$, $\kappa_\text{s}$ and $\kappa_\text{hs}$ are scanned in the interval $(-1,1)$. 

After choosing a set of random input values for the parameters within the regions in Tables \ref{PTtable1} and \ref{PTtable2}, we solve numerically the RGEs given in section \ref{rge}. We repeat numeriously this process by taking input values and solving the RGEs to cover all the allowed regions defined in Tables \ref{PTtable1} and \ref{PTtable2}. Although the initial values we found are suitable at scale $\Lambda\sim 173$ GeV, but as running they vary in the higher scales and may violate one of the conditions e.g. the stability constraint, the positivity condition or the perturbativity. For a random set of inputs we check if all constraints are satisfied up to Planck scale. In the case $m_s>m_H$ the viable initial values which lead to an appropriate result is shown in Fig. \ref{regms>mh} and the viable region of the initial values in the case $m_s<m_H$ is shown in Fig. \ref{regms<mh}. The viable regions given in Figs. \ref{regms>mh} and (\ref{regms<mh}) are for different benchmark singlet scalar VEVs, $w=0.1, 1, 10, 100$ TeV. 

As seen in Fig. \ref{regms>mh} for the case $w=100$ GeV, there is a narrow region which can fulfill the desired condition up to Planck scale. But if we relax the Planck scale implementm then there are a larger viable region for $w=100$ GeV. As we increase the scalar VEV from $w=100$ GeV to $w=100$ TeV, we see in Fig. \ref{regms>mh} that the viable initial values for the couplings $\lambda_\text{s}$ and $\lambda_\text{s}$ shrinks considerably. The coupling $\lambda_\text{s}$ remains $\mathcal{O}(1)$ for all singlet scalar VEV benchmarks. 

In Fig. \ref{regms<mh} that $m_s<m_H$ however the viable region for the case $w=100$ GeV is large and all coupling $\lambda_\text{h}$, $\lambda_\text{s}$ and $\lambda_\text{hs}$ take $\mathcal{O}(1)$ initial values to fulfill the SPP conditions up to Planck scale. As we increase the singlet scalar VEV from $w=100$ GeV to $w=100$ TeV, the couplings $\lambda_\text{s}$ and $\lambda_\text{s}$ must become smaller down to $\mathcal{O}(10^{-8})$ to satisfy the SPP conditions. 
Also in Fig. \ref{runms>mh} for $m_s>m_H$ and in Fig. \ref{runms<mh} for $m_s<m_H$ for each singlet scalar VEV benchmark $w$, we have also shown the evolution of the free parameters for a randomly chosen set of parameters within the regions in Figs \ref{regms>mh} and \ref{regms<mh}. 
The singlet scalar mass is also bounded by the SPP conditions. As seen in Tables \ref{PTtable1} and \ref{PTtable2} the singlet scalar mass varies from about GeV up to about $0.5$ TeV.

\section{Conclusion}\label{concl} \label{conclusion}
The standard Model suffers from a vacuum metastability at high energy scale around $10^{10}$ GeV. The addition of extra scalars in the hidden sector with or without internal symmetries may stablize the vacuum up to the Planck scale. However in general specially when symmetries are absent in the internal configuration space (more investigations needed in future works), the positivity condition might become strong enough to compete with the vacuum stability at a given energy scale. As the simplest example we have investigated a generic real singlet scalar extension of the Standard Model. We have imposed the positivity condition for all scales alongside the absolute vacuum stability and perturbativity to bound the free parameters of the model. As seen in Tables \ref{PTtable1} and \ref{PTtable1}, even before looking at higher scales, the free parameters $\lambda_\text{h},\lambda_\text{s},\lambda_\text{hs},\kappa_\text{s},\kappa_\text{hs}$ at the EW scale are strongly limited due respecting the absolute vacuum stability and the positivity for the vacuum $(v,w)$, with $v=246$ GeV and $w=0.1, 1, 10, 100$ TeV being the Higgs and the singlet scalar vacuum expectation values respectively. The bounds on the free parameter become stronger if we want to keep the vacuum $(v,w)$ to be an absolute minimum, and at the same time respecting the positivity condition and the perturbativity up to the Planck scale (SPP conditions). The result for the viable space of input values for the parameters to respect the SPP conditions is shown in Fig. \ref{regms>mh} when $m_s<m_H$ and in Fig. \ref{regms<mh} when $m_s<m_H$. The upshot is that only for $w=100$ and $m_s<m_H$ all the couplings are $\mathcal{O}(1)$. In other cases, increasing the singlet scale VEV $w$ shrinks the viable region of the couplings down to $\mathcal{O}(10^{-8})$ except for the coupling $\lambda_\text{h}$ which remains in all case $\mathcal{O}(1)$ up to the Planck scale. 

We observe as well that the singlet scalar mass in the presence of the SPP conditions can take values from $\mathcal{O}$(GeV) to $\mathcal{O}$(TeV). 
It is worth to answer this question in future works that how the SPP conditions act on the parameter space of a multi-scalar extension of the SM; whether the SPP condition on multi-scalar theories are more or less restrictive than the simple singlet scalar model we studied here. 

\section*{Acknowledgments}
I would like to thank Alessandro Strumia for useful discussions. This work was supported by the ERC grant NEO-NAT.

\appendix

\bibliography{ref.bib}
\bibliographystyle{plain}

\end{document}